# STAGE STAFFING SCHEME FOR COPYRIGHT PROTECTION IN MULTIMEDIA

Sumit Kumar[1], Santosh Kumar[1], Sukumar Nandi[1]

[1]Department of Computer Science & Engineering
Indian Institute of Technology Guwahati, India
`sumit.itech@gmail.com, santosh.kr@iitg.ernet.in, sukumar@iitg.ernet.in`

## ABSTRACT

*Copyright protection has become a need in today's world. To achieve a secure copyright protection we embedded some information in images and videos and that image or video is called copyright protected. The embedded information can't be detected by human eye but some attacks and operations can tamper that information to breach protection. So in order to find a secure technique of copyright protection, we have analyzed image processing techniques i.e. Spatial Domain (Least Significant Bit (LSB)), Transform Domain (Discrete Cosine Transform (DCT)), Discrete Wavelet Transform (DWT) and there are numerous algorithm for watermarking using them. After having a good understanding of the same we have proposed a novel algorithm named as Stage Staffing Algorithm that generates results with high effectiveness, additionally we can use self extracted-watermark technique to increase the security and automate the process of watermark image. The proposed algorithm provides protection in three stages. We have implemented the algorithm and results of the simulations are shown. The various factors affecting spatial domain watermarking are also discussed.*

## KEYWORDS

*Copyright Protection, Stage staffing watermarking, Digital Multimedia, Encryption and Decryption algorithm for watermarking, Logistic Map, 2-D Arnold Cat Map, Stage Staffing.*

## 1. INTRODUCTION

The enforcement of distribution policies for sensitive intelligence documents is important but difficult. Sensitive documents may be found left behind in conference rooms, common areas, printing rooms, or public folders. Access control based on cryptography alone cannot address this problem. Once after obtaining access to a sensitive document may a person make redundant copies or handle it without care. A major challenge in the reinforcement of sharing policies for sensitive documents is the support of non-repudiation in the primary process so that un-authorized copies of intellect documents can be identified and traced back to their users [1]. The reinforcement should also be appropriate to both hard copies and soft copies of the documents. Conventional cryptographic schemes that cover only soft copies are insufficient to handle this constraint.

Digital watermarking is a promising technology employed by various digital rights management (DRM) systems to achieve rights management [2]. It supports copyright information (such as the owner's identity, transaction dates, and serial numbers) to be embedded as unperceivable signals into digital contents. The signals embedded can be perceivable or insignificant to humans. Literally DRM is a general term used to explain any type of technology that attempts to prevent the piracy. DRM has to tackle with some essential key points to achieve security and fortification in digital multimedia. These essential key points are as following [12]:

- when a document or file can be viewed
- how long a document or file can be viewed for

- whether printing is legitimate and if so how many times
- whether information can be copied and used in other applications
- If a certified user has mislaid their privileges then it ought to be promising to cease their access to information.

The rest of the paper is organized as follow. Section 2 discusses about category and essential properties of digital watermarking. Section 3 explaining implementation scheme and Spread-Spectrum technique of watermarking. We introduce proposed scheme (Stage Staffing Algorithm) in section 4 with detailed analysis of Logistic Map (chaotic map), 2-D Arnold Cat Map Technique and additional enhancement of security using DWT. Section 5 describing simulation results of proposed Stage Staffing Scheme and section 6, 7 are conclusion and future work respectively.

## 2. CATEGORIES AND PERFORMANCE EVALUATION OF DIGITAL WATERMARKING

Basically we can divide water marking techniques in two ways as visible water marking and invisible water marking. Description of them as following as:

### 2.1. Categories of Digital Water Marking

Visible and invisible watermarking are two categories of digital watermarking. The concept of the visible watermark is very simple; it is analogous to stamping a mark on paper. The data is said to be digitally stamped. An example of visible watermarking is seen in television channels when their logo is visibly superimposed in the corner of the screen. Invisible watermarking, on the other hand, is a far more complex concept. It is most often used to identify copyright data such as author, distributor, etc. This paper focuses on this category, and the word "watermark" will mean, by default, the invisible watermark.

In addition, robust watermarking techniques have been designed to oppose tampering and support later extraction and detection of these watermark signals. These signals recover the rights information originally embedded in the document. In this paper we introduce an advanced stage staffing watermarking algorithm, which enables the watermark signal to affectively encrypt into the host image and generate back the watermark. The main idea of algorithm is that there must be a very large difference between the size of host image and the watermark signal. This affects the cyclic behaviour and redundancy in the pixels.

### 2.2. Performance Evaluation of Digital Watermarking

The performance of a watermarking mechanism is evaluated according to: robustness, capacity and imperceptibility. An effective image watermarking algorithm must have the following features:

a) **Imperceptibility**: the watermark must be imperceptible, i.e. the perceived quality of the watermarked image should not be noticeable
b) **Robustness**: The watermark should be difficult, rather impossible, to remove or to degrade, intentionally or unintentionally, by image processing attacks.
c) **Low computational complexity**: The watermarking algorithm should not be computationally complex for embedding as well as extracting the watermark, especially for real time applications.

## 3. PRINCIPLE AND APPLICATION OF WATERMARKING SCHEMES

In this section we present the basic principle of watermarking scheme and describe its practical application. Over the last two decades, digital watermarking has been addressed as an effective solution to safeguard copyright laws and an extensive effort has been made to design robust watermarking algorithms [3].

The two major applications for watermarking are protecting copyrights and authenticating photographs. The main reason for protecting copyrights is to prevent image piracy when the provider distributes the image on the Internet [4]. One way to achieve this goal is by embedding a digital watermark that automatically adjusts itself during image modification [4]. The practice of using images as evidence against crimes, for example, assumes that the images are reliable. Ensuring this requires image authentication [4]. Ensuring the authenticity of an image, i.e., that it has not been tampered with is needed by many organizations, such as hospitals, insurance companies, etc. Many methods are used to authenticate physical images, but this is not the case for digital images. Digital images must be authenticated by digital means. One method authenticates digital images by embedding a digital watermark that breaks or changes as the image is tampered with. This informs the authenticator that the image has been manipulated. In the case of images captured by a digital camera, this can be accomplished by a special chip in the camera, which encrypts a watermark into captured images. This method can also be useful to video and audio formats as well.

### 3.1. Watermarking Insertion Scheme

Digital watermarking describes the process of embedding additional information into a digital media, without compromising the media's value. In this process the embed data called watermark into a media such that watermark can be detected or extracted later to make an assertion about the media [5], [6], [7]. Here input watermark refers to watermark insertion and extract the information is referring as watermark detection. Figure 1 shows the processes for watermark insertion, in which, suppose we have a digital document $X$, a watermark $W$, and a permutation function $\sigma$. A watermark insertion scheme $I$ inserts a watermark $W$ to the document $X$, where $X' = I(X, W, \sigma)$ [7].

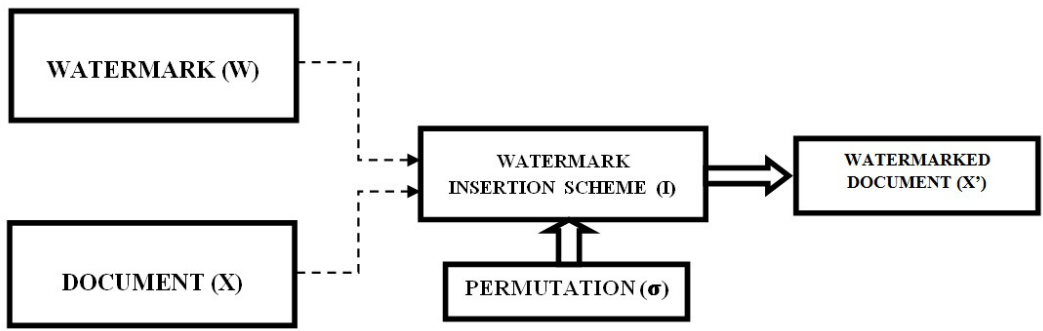

Figure 1. Process of Watermark Insertion

### 3.2. Watermarking Detection scheme

Corresponding to the watermark insertion scheme $I$, there is a watermark detection scheme $D$, which returns a confidence measure of the existence of a watermark $W$ exists in a piece of document $X'$.

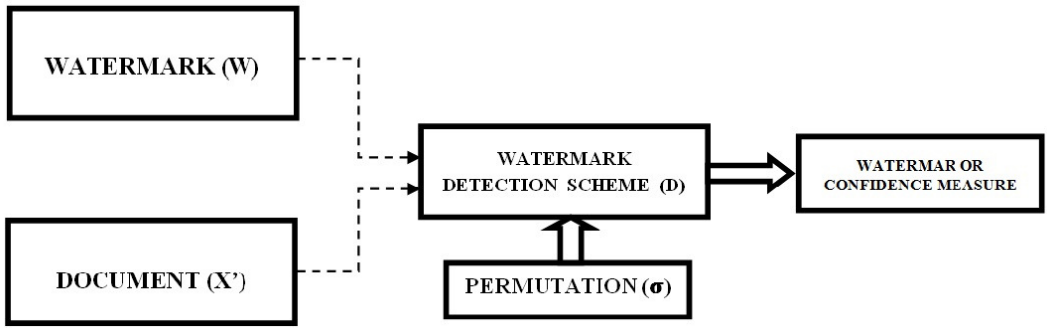

Figure 2. Process of Watermark Detection

Figure 2 Shows the processes for the watermarking detection in which watermark detection scheme D, which returns a confidence measure of the existence of a watermark W exists in a piece of document $X'$. A watermarking technique is referred to as non-blind watermarking when its detection scheme $D$ requires the knowledge of the original document $X$, i.e., $D(X, X', W, \sigma)$ = false if $W$ does not exist in $X'$ or $D(X, X', W, \sigma)$ = true if $W$ exists in $X'$ [7].

## 3.3. The Spread-Spectrum Technique of Watermarking

The spread-spectrum technique assumes:

a) The document is a vector of "features", i.e., $X = \{x_1, x_2, ....., x_n\}$ where $n$ is the number of features in a document.
b) The watermark signal is a vector of "watermark elements", i.e., $W = \{w_1, w_2, ....., x_m\}$ where $m$ is the number of component in a watermark signal, with $n \geq m$.

In which the number of features in a document must be much greater than the number of components in a watermark signal. So the signal is unperceivable in the watermarked document $X'$. The permutation function $\sigma$ is a bisection that shuffles the watermark elements before inserting them to the document $X$. As such, the shuffled watermark is a vector of $\sigma(W) = \{w_1, w_2, ....., x_m\}$ where $w_i = \sigma(w_j)$ with $i, j \leq m$. The permutation function is used for protecting the secrecy of the watermark to be inserted to the document $X$. The shuffled watermark elements are then inserted to the document $X$ by means of a linear insertion operation $\oplus$ (XOR), such that $X'$ in the insertion scheme $I$ is given by:

$$X \oplus \sigma(W) = \{X_1 \oplus W_1', X_2 \oplus W_2', ......, X_n \oplus W_n'\} \quad (1)$$

Spread-Spectrum water marking technique

## 4. PROPOSED WATERMARKING TECHNIQUE

The watermarking used is visually meaningful binary image rather than a randomly generated sequence of bit. Thus human eyes can easily identify the extracted watermark. In fact, embedding a watermark is the least significant bits of pixels are less sensitive to human eyes.

However the watermark will be destroyed if some common image operations such as low-pass filtering are applied to the watermarked image. Therefore, to make the embedded watermark more resistant to any attack, the watermark must be embedded in the more significant bits. This will introduce more distortion to the host image and conflicts with the invisible requirement. To meet both invisible and robust requirement, we will adaptively modify the intensities of some selected pixels as large as possible and this modification is not noticeable to human eyes.

### 4.1. Encryption to Watermark Signal

The watermark image is first converted into the signal form. This can be done in many ways. As converting the signal from binary to ASCII initially, then some sort of encryption is performed. There are many encryption algorithms that completely chaos the watermark signals [8], [9]. We may use 1-D Logistic Map to generate a random sequence. In this, the initial condition and parameters of chaotic map are kept as the secret key; next, the encoded watermark bits are encrypted by chaotic sequence. Therefore, a number of uncorrelated, random like and reproducible encrypted watermarking signals are generated. A commonly used chaotic map is the Logistic map, which is described by equation (2).

$$Z_{n+1} = \mu Z_n (1-Z_n) \qquad (2)$$

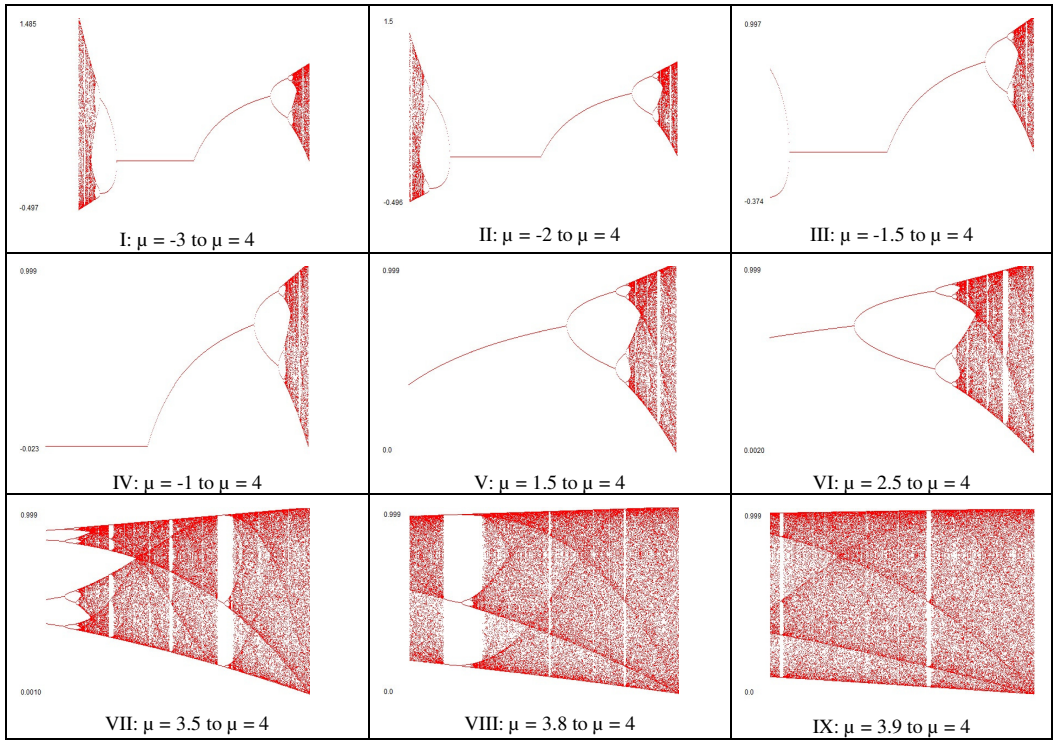

Figure 3. Logistic Map (Chaotic Map)

Where $Z_n$ belongs to $(0,1)$, when $\mu > 3.5699456$, the sequence iterated with an initial value is chaotic, different sequences will be generated with different initial values, as shown in figure 3. We can analyze through figure 3 that end value of μ is fixed and how initial value is generating chaotic effect. The encryption formula is as follows in equation (3).

$$w_{en} = w \oplus c_n \tag{3}$$

Where, $w_{en}$ is the nth encrypted watermark signal, w is the original watermark signal and $c_n$ is the chaotic sequence. The randomizing sequence will give random numbers such that it will be most chaos. But we have assumed a simple DES encryption algorithm to be used as encrypted watermark. This will lead to a sequence of many encrypted watermark signals. Then each sequence is embedded into the Host Image, the image to which copyright protection is needed. The Simple DES Algorithm also provides a good randomness in the watermark signal. This is preferred since, it's easy to implement.

## 4.2. Encryption to the Host Image through Encrypted Watermark Signal

Embedding the watermark signal into the host image is very important task. There are many things to be taken care of while doing this.

a) This must be done in random fashion such that, it is equivalently spread all over.
b) The density of the watermark signal in the host image should be intermediate since, the more the watermarks will be, the more there will be the distortions.

A technique 2-D Arnold Cat Map [9] is used to implement this, as shown in figure 4.

The 2-D Arnold Cat Map is described by equation (4).

$$\left. \begin{array}{l} x_{n+1} = (x_n + y_n) \\ y_{n+1} = (x_n + 2y_n) \end{array} \right\} \bmod 1 \tag{4}$$

Where, notation "$x \bmod 1$" denotes the fractional parts of a real number x by adding or subtracting an appropriate integer. Therefore, $(x_n, y_n)$ is confined in a unit square of $[0,1] \times [0,1]$ we write equation (4) in matrix form to get equation (5).

$$\begin{pmatrix} x_{n+1} \\ y_{n+1} \end{pmatrix} = \begin{pmatrix} 1 & 1 \\ 1 & 2 \end{pmatrix} \begin{pmatrix} x_n \\ y_n \end{pmatrix} = A \begin{pmatrix} x_n \\ y_n \end{pmatrix} \bmod 1 \tag{5}$$

A unit square is first stretched by linear transformation and then folded by modulo operation, so the cat map is area preserving, the determinant of its linear transformation matrix |A| is equal to 1. The map is known to be chaotic. In addition, it is one to one map; each point of the unit square is uniquely mapped onto another point in the unit square. Hence, the watermark pixel of different positions will get a different embedding position. The cat map above can be extended as follows: firstly, the phase space is generalized to $[0,1,2,......,N-1] \times [0,1,2,.......,N-1]$, i.e., only the positive integers from 0 to $N-1$ are taken; and then equation (5) is generalized to give equation (6), 2-D invertible chaotic map.

$$\begin{pmatrix} x_{n+1} \\ y_{n+1} \end{pmatrix} = \begin{pmatrix} a & b \\ c & d \end{pmatrix} \begin{pmatrix} x_n \\ y_n \end{pmatrix} = A \begin{pmatrix} x_n \\ y_n \end{pmatrix} \bmod N \tag{6}$$

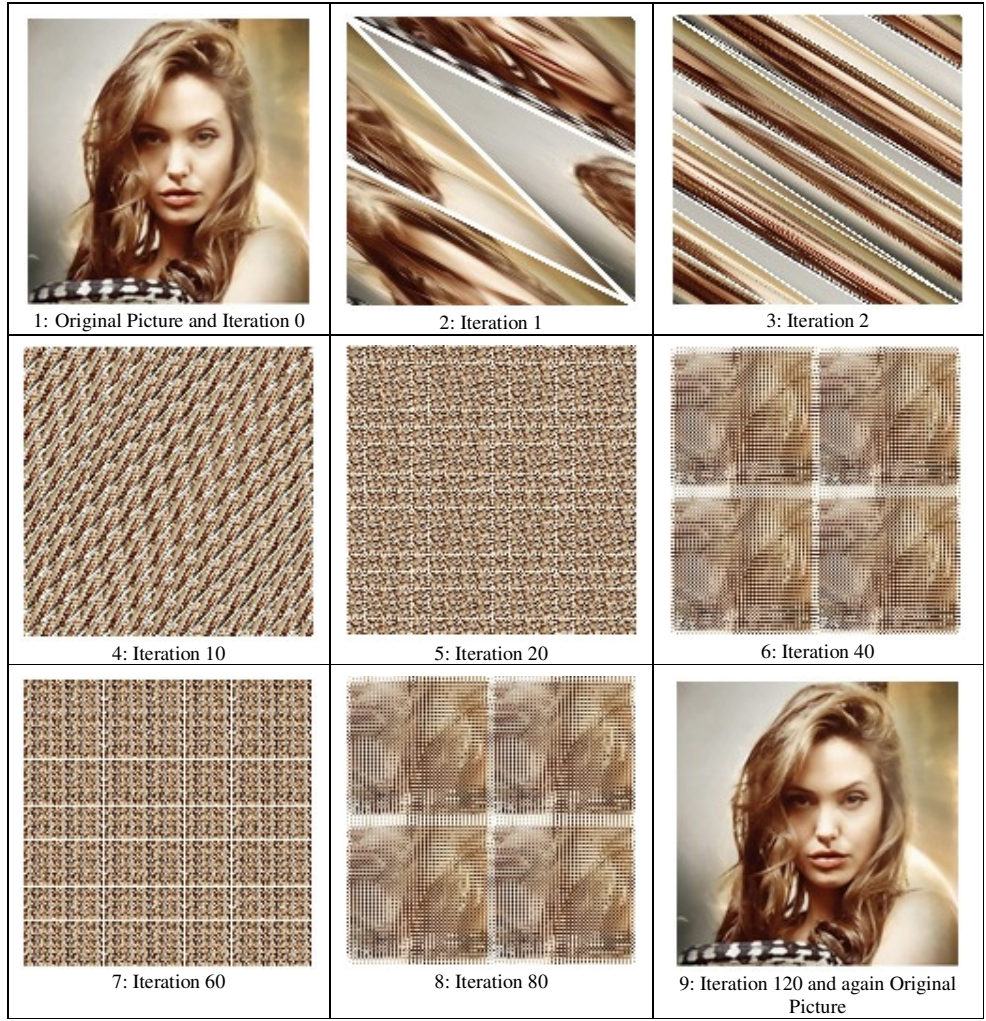

Figure 4. 2-D Arnold Cat Map Technique

Where, a, b, c and d are positive integers, and $|A|= ad - bc = 1$, therefore, only three among the four parameters of a, b, c and d are independent under this condition. The generalized cat map formula is also of chaotic characteristics. By using the generalized cat map, we can obtain the embedding position of watermark pixels, i.e., the coordinate $(i, j)$ of watermark pixel is served as the initial value, three independent parameters and iteration times n are served as the secret key, after n rounds of iterations, the iterating result $(x_n, y_n)$ will be served as the embedding position of watermark pixel of $(i, j)$. When the iteration time's n is big enough, two arbitrary adjacent watermark pixels will separate apart largely in host image; different watermark pixels will get different embedding positions.

But, there is a drawback in this scheme, i.e. if the Watermark image and the host image are of relatively same size then there exists many cycles within the new pixel generation [10]. This is shown in the implementation part. These cycles can't be omitted. Thus, we implement the same as with Stage Staffing watermarking scheme.

## 4.3 Proposed Stage Staffing scheme for Watermarking

In this scheme [11], we create an initial distance such that the watermarked signal equally spreads in the image. Then this distance is iteratively decreased in the quadratic manner and again the encrypted watermark signal is inserted in the image. The number of depth increases the detection of the watermark within the image and comes resistive from the low pass filtering, high pass filtering, median pass filtering attacks etc. The brief algorithm explained by Fig.3.

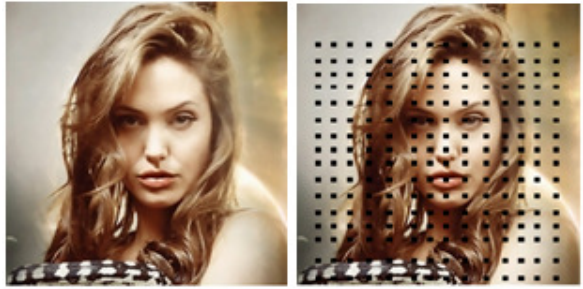

(a) The original host Image (b) First stage of Watermarking

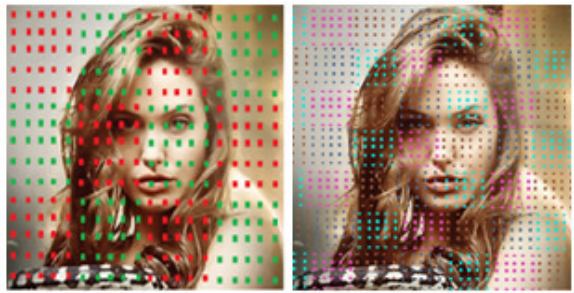

(c) Second stage of Watermarking (d) Third stage of Watermarking

Figure 3. Stages of Stage Staffing scheme for Watermarking

From the above four images, the first image is the original image without any watermark signal embedded into it. The second is the one level watermarking in which the complete image is having one watermark signal with equal distribution. Likewise the third image contains 16 watermark signals. These may be two signals or more than two. The final, the third level is showing the densest watermark in the image. Thus the image consists of 1 + 16 + 64 watermark signals. This gives image high resistivity.

The other important part of this is the place where to embed the watermark signal bit. As we know that for a colour image consists of 24 bits RGB format we can substitute 1 bit in any of these 24 bits. For this, some random number generator algorithm should be used. The bit selected is substituted to 1 if watermark signal is 1 else substituted as 0. If the bit is same previously then no change is made.

Encryption Steps:

1. Generate a watermark signal from the Watermark image.
2. Encrypt this signal to become chaotic. The initial signal and the keys for encryption will be used as secret keys.
3. Embed the encrypt signal in the image in the three stages. i.e., from stage1 to stage3. Usually different bits will be used to store the watermark signal value in the host image.

Extraction Steps:

1. Extract the watermark signals from the image from stage3 to stage1.
2. Decrypt all watermark signals with their respective keys.
3. Average the outcome to give the watermark image.

Hereafter we can add one more step to increase the security of watermark and automate the process of watermark generation using wavelet and wavelet packet 2-D decomposition of image [13]. First is, Self extraction of watermark image is existing method of watermark generation and inclusion of it will dramatically increase the security in copyright protection. Self-extraction of watermark is more secure as compare to normal watermark image, it increases the robustness of protected image, see figure 4 and figure 5.

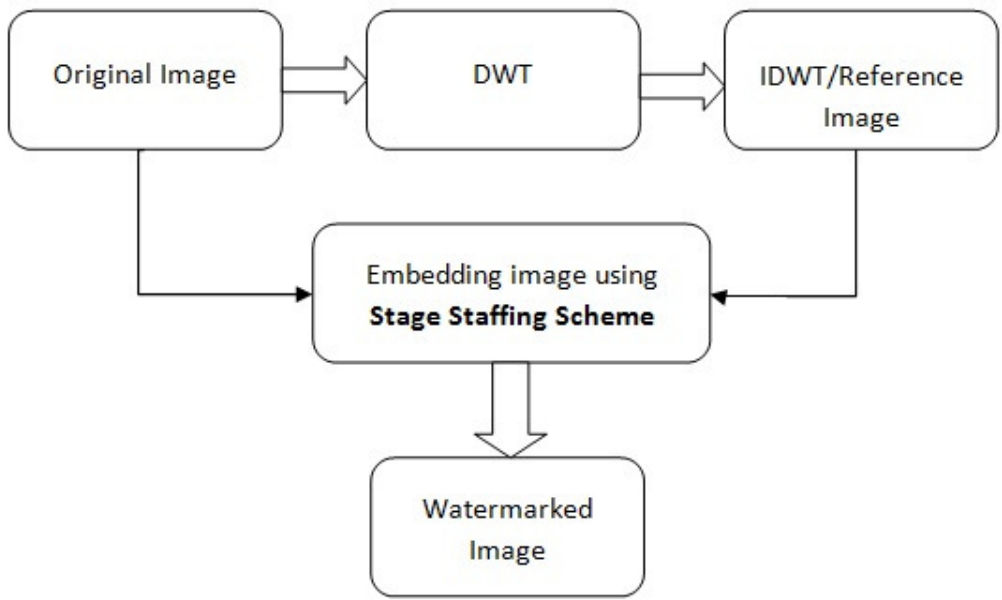

Figure 4. Stage Staffing Scheme using self-extracted watermark using Wavelet decomposition

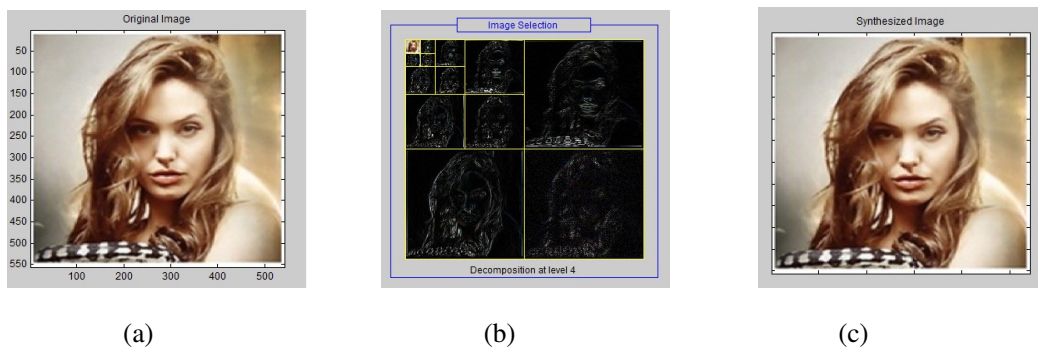

(a)        (b)        (c)

Figure 5. (a)Original Image (b) Decomposition at Level 4 (c) Synthesized Image

Second way of using wavelet decomposition is to apply these decompositions on user defined watermark image. It will increase the chaotic effect and complexity to analyze the copyright method using protected image, see figure 6.

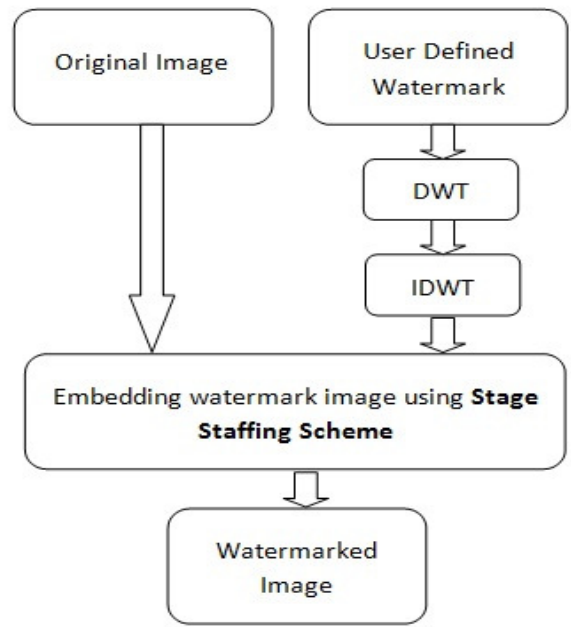

Figure 6. Stage Staffing Scheme using user-defined Wavelet decomposed watermark image

## 5 SIMULATION RESULTS

The algorithm enables the watermark signal to affectively encrypt into the host image and generate back the watermark. From the simulation results, we have observed that there must be a very large difference between the size of host image and the watermark signal. This affects the cyclic behaviour and redundancy in the pixels.

Further, the algorithm can resist through low pass filtering, median pass filtering and high pass filtering. Restriction being that the size of window, used for the filtering must not be large. The regenerated watermark is lost if the window size is taken large.

The stages help to detect watermark in robust conditions. Since it may resist through some noise signals to some extent.

Fig.7. shows various image outputs from the simulation. Description of each is given as follows.

a) The original image used in simulation.
b) The watermark image used for simulation.
c) Image, signal where the watermark is embedded.
d) Regenerated watermark, through extraction procedure.
e) Applying LPF to Watermarked image.
f) Regenerated watermark for LPF.
g) Applying MPF to Watermarked image.
h) Regenerated watermark for MPF.
i) Applying HPF to Watermarked image.
j) Regenerated watermark for HPF.
k) Regenerated watermark signal if the window size is taken large, this is not resistive for this

We are giving here various output of image from simulation result of proposed algorithm.

144

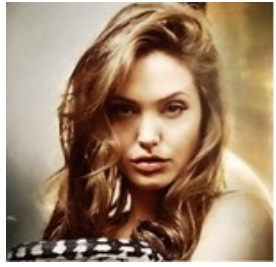 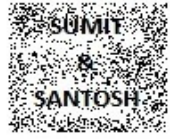

    (a). Original Image         (b). Watermarked Image

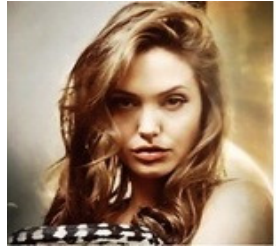 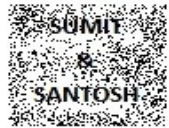

    (c). Encrypted Image       (d). Watermark Regenerated

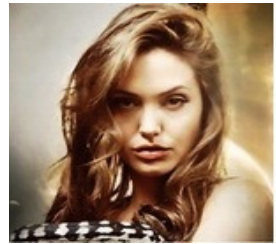 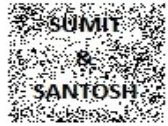

    (e). Image after LPF        (f). Watermark Regenerated

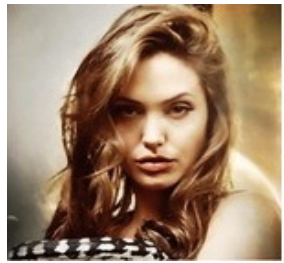 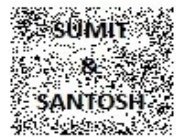

    (g). Image after MPF       (h). Watermark Regenerated

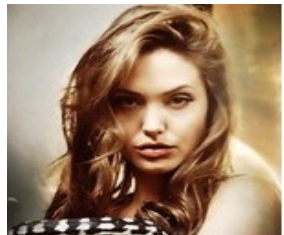 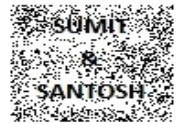

    (i). Image after HPF        (j). Watermark Regenerated

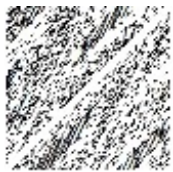

(k). Watermark Regenerated

Figure 7. Enhanced Image and Regenerated Watermark after applying different Image Enhancements.

## 6 CONCLUSIONS

In this paper, a novel watermarking algorithm proposed to address the problem. Proposed algorithm is able to resist attacks of filtering, robustness. A robust watermark scheme based on a block probability for colour image is presented, which operates in spatial domain by embedding the watermark image four times in different positions in order to be robust for cropping attack. The experimental results shows that our scheme is highly robust against various of image processing operations such as, filtering, cropping, scaling, compression, rotation, randomly removal of some rows and columns lines, self-similarity and salt and paper noise. But it is not resistive for several compressions like jpeg etc.

## 7 FUTURE WORK

As we all well-known with thrust of Copyright Protection Schemes to secure digital multimedia, subsequently we are trying to enhance this stage staffing technique with self- referential watermark. In self referential watermarking, watermark will generate form original image with secured chaotic effect in overall process of Copyright Protection.

**Authors**

**Mr. Sumit Kumar** received M.Tech (Computer Science and Engineering) degree from Indian Institute of Technology Guwahati then He joined Indian Institute of Technology Patna as Research Fellow in Department of Computer Science and Engineering.

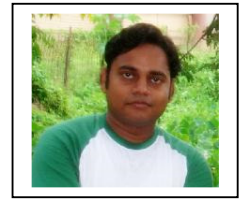

**Mr. Santosh Kumar** received M.Tech (Computer Science and Engineering) degree from Indian Institute of Technology Guwahati then he joined Indian Institute of Technology Guwahati as Junior Project Fellow in Department of Computer Science and Engineering.

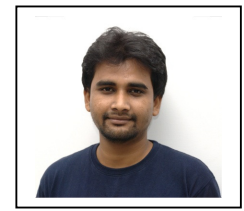

**Dr. Sukumar Nandi** received B Sc (Physics), B Tech and M Tech from Calcutta University in 1984, 1987 and 1989 respectively.He received the Ph D degree in Computer Science and Engineering from Indian Institute of Technology Kharagpur in 1995. In 1989-90 he was a faculty in Birla Institute of Technology, Mesra, Ranchi, India. During 1991 to 1995, he was a scientific officer in Computer Sc & Engg, Indian Institute of Technology Kharagpur.In 1995 he joined in Indian Institute of Technology Guwahati as an Assistant Professor in Computer Science and Engineering.Subsequently, he became

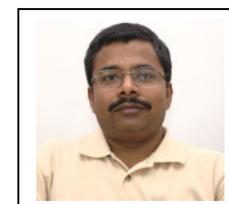

Associate Professor in 1998 and Professor in 2002. He was in School of Computer Engineering, Nanyang Technological University, Singapore as Visiting Senior Fellow for one year (2002-2003). He was member of Board of Governor, Indian Institute of Technology Guwahati for 2005 and 2006. He was General Vice-Chair of 8th International Conference on Distributed Computing and Networking 2006. He was General Co-Chair of the 15th International Conference on Advance Computing and Communication 2007. He is also involved in several international conferences as member of advisory board/ Technical Programme Committee. He is reviewer of several international journals and conferences. He is co-author of a book titled "Theory and Application of Cellular Automata" published by IEEE Computer Society. He has published more than 150 Journals/Conferences papers. His research interests are Computer Networks (Traffic Engineering, Wireless Networks), Computer and Network security and Data mining. He is Senior Member of IEEE and Fellow of the Institution of Engineers (India).